\documentclass[a4paper,11pt]{article}
\pdfoutput=1 
\usepackage[export]{adjustbox}

\usepackage{jcappub}
\usepackage{graphicx}
\usepackage{dcolumn}
\usepackage{amssymb,amsmath,bm}
\usepackage{mathtools}
\usepackage{color}
\usepackage[dvipsnames]{xcolor}
\usepackage{xfrac}
\usepackage{aas_macros}
\usepackage{mathrsfs}
\usepackage{subcaption}
\usepackage{rotating}
\usepackage{chngcntr}
\usepackage[T1]{fontenc} 
\usepackage{natbib}
\usepackage{lipsum}
\usepackage{hyperref}
\newcommand{\nv}{\hat{\bm n}}

\newcommand{\deq}{\coloneqq}

\newcommand{\Ecut}{E_\text{cut}}
\newcommand{\Ecuti}[1]{E_{\text{cut},\,#1}}
\newcommand{\cN}{\mathcal{N}}
\newcommand{\cS}{\mathcal{S}}

\newcommand{\de}{\mathrm{d}}

\newcommand{\ra}{\rightarrow}
\newcommand{\nn}{\nonumber}

\newcommand{\fp}{f_\text{p}}
\newcommand{\fH}{f_\text{H}}
\newcommand{\fFe}{f_\text{Fe}}

\newcommand{\fmix}{f_\text{mix}}
\newcommand{\nev}{N_\text{ev}}

\def\dd{ \text{d} }

\def\+{\dagger}
\def\la{\langle}
\def\ra{\rangle}

\def\n0{{\la n \ra}}

\def\B0{{\la B \ra}}

\title{Constraining ultra-high-energy cosmic ray composition through cross-correlations}
\author[a,\,1]{Konstantinos Tanidis,}
\author[a]{Federico R.~Urban,}
\author[b,\,c,\,d]{and Stefano Camera}

\affiliation[a]{CEICO, Institute of Physics of the Czech Academy of Sciences,\\Na Slovance 1999/2, 182 21 Prague, Czech Republic}
\affiliation[b]{Dipartimento di Fisica, Universit\`a degli Studi di Torino,\\Via P.\ Giuria 1, 10125 Torino, Italy}
\affiliation[c]{INFN -- Istituto Nazionale di Fisica Nucleare, Sezione di Torino,\\Via P.\ Giuria 1, 10125 Torino, Italy}
\affiliation[d]{INAF -- Istituto Nazionale di Astrofisica, Osservatorio Astrofisico di Torino,\\Strada Osservatorio 20, 10025 Pino Torinese, Italy}

\footnotetext[1]{email: tanidis@fzu.cz}

\abstract{The chemical composition of the highest end of the ultra-high-energy cosmic ray spectrum is very hard to measure experimentally, and to this day it remains mostly unknown. Since the trajectories of ultra-high-energy cosmic rays are deflected in the magnetic field of the Galaxy by an angle that depends on their atomic number \(Z\), it could be possible to indirectly measure \(Z\) by quantifying the amount of such magnetic deflections.  In this paper we show that, using the angular harmonic cross-correlation between ultra-high-energy cosmic rays and galaxies, we could effectively distinguish different atomic numbers with current data. As an example, we show how, if \(Z=1\), the cross-correlation can exclude a \(39\%\) fraction of Fe56 nuclei at \(2\sigma\) for rays above \(100~\text{EeV}\).}

\begin{document}
\maketitle
\flushbottom

\section{Introduction}\label{sec:intro}

Sixty-one years since the discovery of ultra-high-energy cosmic rays (UHECRs) have passed~\cite{Linsley:1961kt}, and, despite extensive experimental efforts, we still do not know what the UHECRs at the high end of the energy spectrum are~\cite{AlvesBatista:2019tlv}. More precisely, we know that these UHECRs are charged nuclei---neutral particles can make up only a marginal fraction of the detected UHECR flux at the highest energies~(see e.g.\ \cite{TelescopeArray:2021fpj})---but direct measurements of the UHECR atomic number(s) \(Z\) (i.e., their chemical composition) are, to this day, inconclusive~\cite{AlvesBatista:2019tlv,Zhezher:2021qke,PierreAuger:2021mmt}. This is because these measurements are prone to systematics (e.g.\ the extrapolation of low-energy hadronic models to much higher energies) and because at the highest energies the flux of UHECRs is extremely low.

Indirect information about the atomic number(s) \(Z\) of UHECRs can be obtained from the UHECR energy spectrum and their arrival direction distribution. In order to exploit these observables one needs to model, i.e., make assumptions, about (a) the distribution of the sources of UHECRs in the sky and (b) the propagation of UHECRs from the sources to the Earth. As for the distribution and properties of sources, even though the details remain unclear, recent experimental data point towards the so-called large-scale structure (LSS) model, in which the UHECR sources are numerous, steady, and follow closely the distribution of matter in the local Universe~\cite{Ding:2021emg} (however see~\cite{TelescopeArray:2021gxg,Allard:2021ioh}). We adopt this model in this study (see \autoref{sec:modelling} for details).

The propagation of UHECRs from source to Earth is determined by two distinct effects. The first is the energy losses of UHECRs, which is caused by the interactions of UHECRs with a cosmological background of low-energy photons and by the expansion of the Universe. This effect can be calculated thanks to state-of-the-art propagation codes such as \emph{SimProp}~\cite{Aloisio:2017iyh} once the injection model, that is, the atomic number and the injection energy power spectrum at the source, is specified. The second effect is the deflection of UHECRs in intervening magnetic fields, most importantly the Galactic magnetic field (GMF)~\cite{Haverkorn:2014jka,Boulanger:2018zrk}.\footnote{If cosmological fields of magnitude \(B\sim10^{-12}~\text{G}\), and not more, are the precursors to the fields we observe today in galaxies and clusters of galaxies, extra-Galactic fields play a very small role in UHECR propagation~\cite{Dolag:2004kp}. Although this is far from being settled, we choose to neglect hypothetical cosmological magnetic fields in this study.}

Among these effects, the most informative about the atomic number \(Z\) is the deflection of UHECR arrival directions. This is the case because the angle \(\theta\) by which a charged particle is deflected in the ballistic regime (which applies for the energies \(E\) and magnetic field values \(B\) we consider in this work), scales as \(\theta\propto Z B/E\), and the atomic number \(Z\) is the least-known quantity in this expression. Indeed, the energy \(E\) of a UHECR is determined with about 20\% accuracy from experimental data~\cite{TelescopeArray:2021zox}; the GMF strength \(B\) is of order of a few \(\mu\mathrm{G}\) of magnitude, and, although its variation across the Galaxy is not well known, the uncertainty is limited to within a factor of a few even in the most complex and hard-to-test regions~\cite{Haverkorn:2014jka,Boulanger:2018zrk}.\footnote{It is also possible to obtain some nearly-model-independent information about UHECR deflections in the GMF from Faraday rotation measures data, once the atomic number is specified~\cite{Pshirkov:2013wka}} The atomic number instead ranges from \(Z=1\) for protons (H1 nuclei) to \(Z=26\) for iron (Fe56 nuclei), and is the most relevant unknown parameter that shapes the UHECR arrival distribution in the sky.

Taking stock of this observation, in this work we build a tool which can contribute to determine (or exclude) atomic number models in as much as a model-independent way as possible. This approach is of course not new, and was most recently discussed in~\cite{Ahlers:2017wpb,dosAnjos:2018ind,Kuznetsov:2020hso}.\footnote{Several other studies have used various combinations of data to place theoretical or experimental constraints on the atomic number, but in a more indirect, or ``holistic'' way, see~\cite{Kuznetsov:2020hso} and references therein.} Our contribution here is to focus on a set of test statistics (TS) that are derived from the harmonic decomposition of the UHECR flux across sky directions \(\nv\) as \(\Phi(\nv) = \sum a_{\ell m} Y_\ell^m(\nv)\) where the \(a_{\ell m}\) coefficients quantify the anisotropy on angular scales of order \(\pi/\ell\) and \(Y_\ell^m(\nv)\) are spherical harmonics. More precisely, in this paper we assess how well a full-sky UHECR experiment with statistics comparable to current facilities can discriminate among atomic numbers, using TSs built from the harmonic angular UHECR auto-correlation (AC) and UHECR-galaxies cross-correlation (XC) power spectra. The AC and XC have been already recognised as useful tools to detect the angular anisotropy of the UHECR flux~\cite{Sommers:2000us,Tinyakov:2014fwa,Deligny:icrc2015,diMatteo:2018vmr,Urban:2020szk}.

Our main result is that, with current data, the XC should moderately outperform the AC in constraining the atomic number \(Z\) of UHECRs. This remains true for different choices of UHECR energy cuts, fiducial model and TS. We find that the most informative TS is the total harmonic angular power, which can more easily distinguish different atomic numbers \(Z\) compared to individual multipoles such as the harmonic dipole or quadrupole.

This paper is structured as follows. In the following \autoref{sec:modelling} we introduce our models for the UHECR sources and their propagation, including intervening magnetic fields. In \autoref{sec:method} we review the properties of the AC and XC, explain our choice of TSs and outline the procedure to obtain our forecasts. In \autoref{sec:results} we show the main results of our simulations, which we discuss in \autoref{sec:conclusions} where we also give an outlook for future work. Appendix~\ref{app:some} collects additional results which are provided for reference.

\section{Modelling}\label{sec:modelling}

In this section we describe our models for the injection properties of UHECRs at the sources (\autoref{ssec:uhecr}), their displacement by way of the GMF on their way to the Earth (\autoref{ssec:gmf}), the observed energy spectrum of UHECRs (\autoref{ssec:spec}) and the galaxy catalogue that we use as a proxy for the cosmological distribution of UHECR sources (\autoref{ssec:gal}).

\subsection{Cosmic ray injection}\label{ssec:uhecr}
Our main working assumptions are that UHECR sources are numerous (more than \(1\,\mathrm{Mpc}^{-3}\) \citep{Waxman:1996hp}), steady, and that their distribution is correlated with the distribution of galaxies. Specifically, we assume that each source injects a flux of UHECRs with a power-law spectrum \(\varphi\propto E^{-\gamma}\) up to a very high energy. Therefore, the integral UHECR flux above a certain energy cut \(\Ecut\), in a given direction \(\nv\), and for a single species with atomic number \(Z\), reads
\begin{align}\label{eq:cr_flux}
	\Phi(\Ecut,\nv;\gamma,Z)& \deq \frac{\mathcal{E}_0}{4\pi}\,\frac{\bar{n}_{\rm s,c}\,\Ecut}{\gamma-1}\left(\frac{\Ecut}{E_0}\right)^{-\gamma} \int \de\chi \frac{\alpha(z,\Ecut;\gamma,Z)}{(1+z)}\,\left[1+\delta_{\rm s}(z,\chi\nv)\right] \,,
\end{align}
where \(\bar{n}_{\rm s,c}\) is the comoving average number density of sources (which is assumed not to be evolving with redshift), \(\mathcal{E}_0\) is the overall emissivity normalisation factor, \(E_0\) the energy at which the flux is normalised, \(z\) is cosmological redshift, \(\chi\) is the radial comoving distance, such that \(\de\chi/\de z=1/H(z)\), with \(H\) the Hubble factor, and \(\delta_{\rm s}(z,\chi\nv)\) is the UHECR source density contrast (for a derivation and further references see~\citep{Urban:2020szk}). The function \(\alpha(z,\Ecut;\gamma,Z)\) is the attenuation function that gives the probability that a UHECR detected with energy above \(\Ecut\) had originated from a source located at redshift \(z\) and which emits UHECRs with atomic number \(Z\) and injection slope \(\gamma\). We calculated the attenuation function by following \(10^6\) UHECRs with \emph{SimProp} v2r4~\cite{Aloisio:2017iyh} with energies above \(E = 10~\text{EeV}\) (with an upper cutoff of \(E_\text{max} = 10^5~\text{EeV}\)), for redshifts up to \(z = 0.3\). With \emph{SimProp} we accounted for all energy losses, namely adiabatic losses and losses due to interactions with cosmic microwave background photons and with extra-Galactic background photons according to the model in Ref.~\citep{Stecker:2005qs}.

We are interested in the anisotropies in the UHECR flux, namely
\begin{align}\label{eq:cr_ani}
    \Delta(\Ecut,\nv;\gamma,Z) &\deq \frac{\Phi(\Ecut,\nv;\gamma,Z)}{\bar{\Phi}(\Ecut;\gamma,Z)}-1= \int\dd\chi\,\phi(\Ecut,\chi;\gamma,Z)\,\delta_{\rm s}(z,\chi\nv) \,,
\end{align}
where \(4\pi\bar{\Phi}_Z(\Ecut)\deq \int\dd\nv\,\Phi(\Ecut,\nv;\gamma,Z)\) is the average flux across the sky. In the second definition we have recast the anisotropy as an integral of the UHECR radial kernel
\begin{align}\label{eq:cr_ker}
    \phi(\Ecut,z;\gamma,Z) &\deq \left[\int\dd \tilde{z}\,\frac{\alpha(\tilde{z},\Ecut;\gamma,Z)}{H(\tilde{z})(1+\tilde{z})}\right]^{-1} \frac{\alpha(z,\Ecut;\gamma,Z)}{(1+z)} \,,
\end{align}
where \(H(z)\) is the Hubble parameter.

In order to best illustrate our method we adopt a similar strategy as~\cite{Kuznetsov:2020hso}, and inject a variable mixture of two primaries, with the fraction of the heavier nucleus \(\fmix\) being the free parameter that we seek to determine or constrain. This can be generalised for any injection model that can be parametrised by a set of fractions that determine the relative weights of each element. The radial kernel for the admixture of the two species \(Z_1\) and \(Z_2\) becomes
\begin{align}\label{eq:cr_mix}
    \phi_\text{mix}(\Ecut,z;\gamma,Z_1,Z_2) &\propto \frac{(1-\fmix)\,\alpha(z,\Ecut;\gamma,Z_1)+\fmix\,\alpha(z,\Ecut;\gamma,Z_2)}{(1+z)} \,.
\end{align}
In principle we can generalise the mixed kernel \autoref{eq:cr_mix} to the case of different injection slopes \(\gamma_1\) and \(\gamma_2\).

Heavy nuclei with energy \(E\gg A\,\Ecut\), with \(A\) their mass number, disintegrate rapidly on their way from the source to the Earth. This means that the attenuation function \(\alpha(z,\Ecut;\gamma,Z)\) for \(Z>1\) is going to be broken up into several attenuation functions, one for each species that the original nucleus breaks into. For simplicity we follow~\cite{diMatteo:2017dtg} and approximate the flux coming from a heavy \((A,Z)\) nucleus as the sum of a fraction \(\fp = A^{2-\gamma}/(A^{2-\gamma}+1)\) from ``light'' elements (which we count as protons), and the remaining \(1-\fp\) part from ``heavy'' elements (which we count as the original nuclei)---notice that we can still use this formula even for \(A=1\), although physically nothing happens in this case. This choice is further justified because the uncertainties on the magnetic deflections and injection spectrum have a far larger impact on the harmonic correlations than the error that we introduce by ignoring the details of the mass composition. Therefore, in this approximation we substitute
\begin{align}\label{eq:att_p}
    \alpha(z,\Ecut;\gamma,Z) &\rightarrow \fp\,\alpha(z,\Ecut;\gamma,1) + (1-\fp)\,\alpha(z,\Ecut;\gamma,Z) \,.
\end{align}
For instance, for \((A,Z)=(56,26)\), that is, Fe56, and an injection slope of \(\gamma=2.3\) we find \(\fp \approx 0.23\), which is a significant fraction. Conversely, for energies \(E\ll A\Ecut\) nearly all nuclei that reach the Earth with \(E\geq\Ecut\) remain intact.

\subsection{Magnetic fields}\label{ssec:gmf}

Upon reaching the Milky Way the UHECRs meet the GMF screen, which deviates their trajectories on their way to the Earth and obfuscates the original anisotropy of their arrival direction distribution. The GMF has an amplitude of about a few \(\mu\mathrm{G}\), and a complex three-dimensional structure, which to date is still poorly understood~\cite{Haverkorn:2014jka,Boulanger:2018zrk,Unger:2017kfh}; nonetheless, some quantitative statements can be made, and as a reference we expect that protons with energy \(E=100\,\mathrm{EeV}\) will be deflected by a few degrees for the most part of the sky, except for certain directions close to the Galactic plane. In a simplified treatment we can account for the effects of the GMF by smearing the map of sources below a certain angular scale. The magnetic beam reads
\begin{align}\label{eq:beam}
    {\cal B}(r) &\deq \frac{1}{2\pi\sigma^2}\exp\left[-\frac{r^2}{2\sigma^2}\right] \,,
\end{align}
where \(r=|\nv_1-\nv_2|\) for two directions \(\nv_1\) and  \(\nv_2\), and the width of the Gaussian beam, the displacement \(\sigma\),\footnote{The deflection angle \(\theta\), which is the angle by which a charged particle is deflected from its trajectory in a magnetic field, is not the same as the observable displacement angle \(\sigma\), which is the angle in the sky between the original and actual directions of the charged particle. The two quantities are related as \(\langle\sigma^2\rangle=\langle\theta^2\rangle/3\)~\cite{Harari:2002dy}} is expressed as
\begin{align}\label{eq:smear}
    \sigma \deq \frac{1}{\sqrt2}\left(\frac{40\,\text{EeV}}{E/Z}\right) \frac{1\,\text{deg}}{\sin^2 b + 0.15} \,,
\end{align}
where \(b\) is galactic latitude, see~\cite{Pshirkov:2013wka,diMatteo:2017dtg} (notice the factor \(1/\sqrt2\) difference with the definition and normalisation of these works).

In order to be able to treat this problem analytically we conservatively smear uniformly across the sky with the maximum value of \(\sigma\), obtained from \(b=0\); this gives a displacement of approximately 4.7~deg for 40~EeV protons.\footnote{Although this formula does not strictly extend to the much larger displacements that would be experienced by e.g.\ Fe56, we retain it as a toy model to illustrate our method, see \citep{upcoming} for further details.} The possibility of an extra-Galactic magnetic field can be accounted for, in our formalism, in the same way, increasing \(\sigma\) accordingly. In order to illustrate how our results depend on the choice of magnetic smearing, in \autoref{app:deflect} we present results for a half and twice as much deflections as in our reference model, \autoref{eq:smear}. This approach does not take into account the structure of the large-scale GMF, whose magnification and demagnification of the sources induce significant distortions on the UHECR anisotropy patterns. However, the harmonic angular power spectra, being global measures of the anisotropy, are less affected by coherent deflections than positional quantities.

Because the magnetic deflections depend not only on \(Z\) but also on the energy, we can improve on the analysis by binning the UHECR flux in five logarithmic energy bins with logarithmic width of 0.1, and compute the total flux by adding them up proportionally to the UHECR energy spectrum (see below)---we do this for each UHECR species separately. 
The kernel \autoref{eq:cr_ker} is generalised as
\begin{align}\label{eq:cr_bin}
	\phi(\Ecut,z;\gamma,Z)&\propto \sum_i Q_i \, \frac{\alpha(z,\Ecuti{i}\,;\gamma,Z)}{1+z} \,,
\end{align}
where \(Q_i\) are the fractions of flux in the energy bin \(i\) for which \(\Ecut\in[\Ecuti{i}\,,\Ecuti{i+1}]\) (for the last bin we set \(\Ecuti{i+1}=\infty\)), as obtained from the energy spectrum \autoref{eq:en_spectrum} below. We then apply to each bin a magnetic smearing obtained from \(\Ecuti{i}\).
 
\subsection{Energy spectrum}\label{ssec:spec}

The energy spectrum on Earth, and with it the number of UHECR events, should be self-consistently derived from the injection spectrum propagated from the source. However, within our level of accuracy, we can more simply choose the final energy spectrum in such a way that it is representative of what is observed in current experimental facilities. Therefore, in order to determine the number of UHECR events that in turn fix the levels of Poisson noise in the angular power spectra, we define the differential observed energy spectrum as
\begin{align}\label{eq:en_spectrum}
    J(E)=&  J_0 \left(\frac{E}{\text{EeV}}\right)^{-\gamma_1} \quad E \leq E_1 \nn\\
    J(E)=&  J_0 \left(\frac{E_1}{\text{EeV}}\right)^{-\gamma_1}\left(\frac{E}{E_1}\right)^{-\gamma_2} \quad E > E_1 \,,
\end{align}
where \(\gamma_1 = 3\), \(\gamma_2 = 5\), \(E_1 = 10^{19.75}~\text{eV}\), and \(J_0 = 4.28\times10^6/\text{EeV}\) as benchmark values. This choice gives a total number of events above \(\Ecut\), defined as \(\nev\deq \int_{\Ecut}\dd E\,J(E)\), of approximately 1000, 200, 30 events for \(\Ecut=10^{19.6}\,\mathrm{eV}\approx40\,\mathrm{EeV}\), \(\Ecut=10^{19.8}\,\mathrm{eV}\approx63\,\mathrm{EeV}\), \(\Ecut=10^{20}\,\mathrm{eV}=100\,\mathrm{EeV}\), respectively. These are the number of events we can expect to have for a full-sky observatory with statistics comparable with current experimental facilities~\cite{Ivanov:2021mkn,PierreAuger:2021ibw}. For each \(\Ecut\) we compute the corresponding attenuation and define five logarithmic bins as described in \autoref{ssec:gmf} for the deflections, analogously to the UHECR kernel \autoref{eq:cr_bin}. 


\subsection{Galaxies}\label{ssec:gal}

The galaxy sample anisotropy, using the same language, is defined as
\begin{align}\label{eq:g_ani}
    \Delta_\text{g}(\nv) &\deq \frac{N_\text{g}(\nv)}{\bar{N}_\text{g}}-1 =\int \de\chi\;\phi_\text{g}(\chi)\,\delta_\text{g}(z,\chi\,\nv)\,,
\end{align}
where \(N_\text{g}(\nv)\) is the number of galaxies in a given direction \(\nv\) and \(\bar{N}_\text{g}\) its average over the observed patch of sky. Analogously to the UHECR anisotropy \autoref{eq:cr_ani}, we define the three-dimensional galaxy overdensity as \(\delta_\text{g}(z,\chi\,\nv)\), and \(\phi_\text{g}(\chi)\) is kernel of the galaxy distribution, i.e.\ the weighted distribution of galaxy distances (see \autoref{eq:cr_ker}). The galaxy kernel \(\phi_\text{g}(\chi)\) is given by
\begin{align}\label{eq:g_ker}
	\phi_\text{g}(\chi)\deq \left[\int \de\tilde\chi\; \tilde\chi^2\,w(\tilde\chi)\,\bar{n}_{\rm g,c}(\tilde\chi)\right]^{-1}\,\chi^2\,w(\chi)\,\bar{n}_{\rm g,c}(\chi) \,,
\end{align}
where \(\bar{n}_{\rm g,c}(\chi)=\chi^{-2}\de N_\text{g}/\de\chi\) is the comoving, volumetric number density of galaxies in the sample with \(\de N_\text{g}\) the number of galaxies in a bin of width \(\de\chi\). The quantity \(w(\chi)\) is an optional distance-dependent weight that can be applied to all the objects in the galaxy catalogue, provided their redshifts are known. Assuming Poisson statistics, it can be shown~\cite{Alonso:2020mva} that the optimal weights that maximise the signal-to-noise ratio of the galaxy-UHECR XC are given by
\begin{align}\label{eq:opt_w}
    w(\chi) &= \frac{\alpha(z,\Ecut;\gamma,Z)}{(1+z)\chi^2\,\bar{n}_{\rm g,c}(\chi)} \,.
\end{align}
The optimal weights \autoref{eq:opt_w} are equal to the UHECR kernel divided by the galaxy kernel. Physically this means that the best signal-to-noise ratio is obtained by up-weighing galaxies with low redshift and down-weighing galaxies at high redshift, in such a way that their weighted distribution tracks the probability of UHECRs to come from a given redshift (or distance).

\begin{figure}[htbp]
    \centering
    \includegraphics[width=1.0\columnwidth]{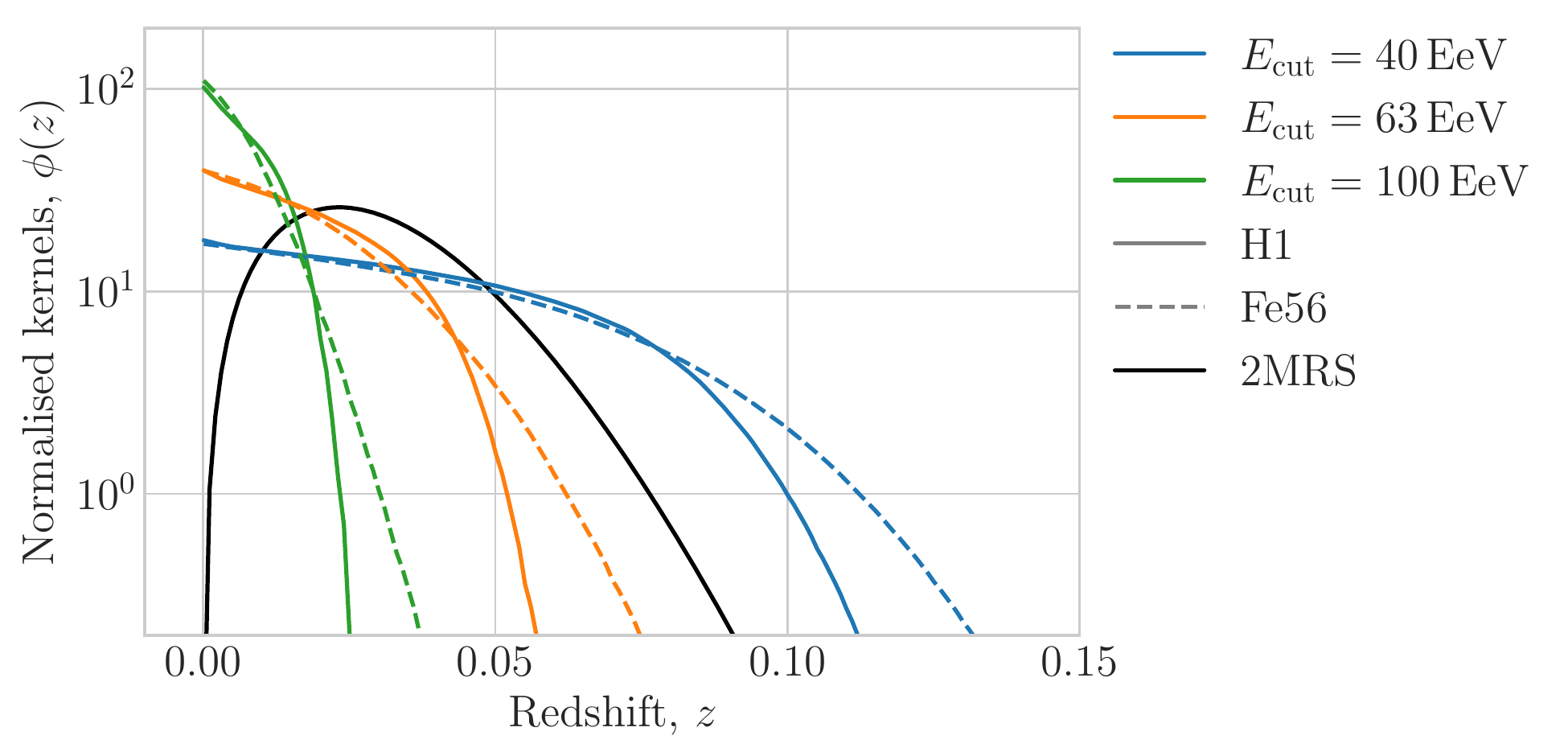}
    \caption{Normalised kernels for (unweighted) galaxies and UHECRs at different energy cuts, in black or colour respectively. Note that all kernels are normalised to unit area.}
    \label{fig:Nz}
\end{figure}

The properties of the galaxy sample are modelled after the 2MASS Redshift Survey (2MRS)~\cite{Huchra:2011ii}, which constitutes one of the most complete full-sky spectroscopic low-redshift surveys, generalised to the full sky. The 70\% sky coverage for the actual 2MRS catalogue would only degrade our signal by a factor of \(\sqrt{0.7}\); for simplicity we also assume that the catalogue is complete. 
The redshift distribution of galaxies is well fitted by
\begin{align}\label{eq:2MRS}
    \frac{\de N_g}{\de z}=\frac{\bar N_g\,\beta}{z_0\,\Gamma\left[(m+1)/\beta\right]}\left(\frac{z}{z_0}\right)^m \exp{\left[-{\left(\frac{z}{z_0}\right)}^\beta\right]}
\end{align}
with \(\bar N_g=43\,182\) the total number of sources, \(\beta=1.64\), \(m=1.31\) and \(z_0=0.0266\) (see \citep{Ando:2017wff}). \autoref{fig:Nz} shows in black the galaxy kernel as given by \autoref{eq:2MRS}, normalised to unit area.

Given that the number of detected UHECRs is small (for \(E\gtrsim40~\text{EeV}\) it is about 1000) we assume that on average each UHECR comes from a different source, which emits UHECRs isotropically, and that all UHECR sources are in the galaxy catalogue, such that \(\delta_\text{g}(z,\chi\,\nv) = \delta_{\rm s}(z,\chi\,\nv)\).\footnote{Notice that the overdensities themselves are never needed, only their two-point correlators \(\la\delta_a\delta_b\ra\) are. The correlators are proportional to the Fourier power spectrum \(P){ab}(k)\) as defined in the appendix of~\cite{Urban:2020szk}, where the relationship to the overdensity is also given explicitly---see also \autoref{sec:method}.} This assumption is accurate if the sources are dense enough with respect to the number of UHECR events \citep{Koers:2008ba}, as is the case of all galaxy surveys for cosmology, like the one adopted here.

\section{Method}\label{sec:method}

The two-dimensional fields \(\Delta_a(\nv)\) can be decomposed into their harmonic coefficients
\begin{align}\label{eq:sph_harm}
	\Delta_{\ell m}^a &\deq \int\dd\nv\,Y^*_{\ell m}(\nv)\,\Delta_a(\nv) \,,
\end{align}
where \(Y_{\ell m}\) are Laplace's spherical harmonics and \(a\in\left\{\text{CR},\,\text{g}\right\}\) for UHECRs and galaxies, respectively. The covariance of the \(\Delta_{\ell m}\) is the signal angular power spectrum \(\cS^{ab}_\ell\), defined as
\begin{align}\label{eq:aps}
	\langle \Delta^a_{\ell m}\,\Delta^{b\ast}_{\ell'm'}\rangle &\deq \delta_{\ell\ell'}\,\delta_{mm'}\,\cS^{ab}_\ell \,.
\end{align}

For broad kernels such as \autoref{eq:cr_ker} and \autoref{eq:g_ker} the harmonic angular power spectrum \(\cS_\ell^{ab}\) between two projected quantities \(\Delta_a\) and \(\Delta_b\) is related to their three-dimensional Fourier-space power spectrum \(P_{ab}(z,k)\) by
\begin{equation}\label{eq:cl_limber}
	\cS^{ab}_\ell=\int \frac{\de\chi}{\chi^2}\;\phi_a(\chi)\,\phi_b(\chi)\,P_{ab}\left[z(\chi),k=\frac{\ell+1/2}{\chi}\right] \,,
\end{equation}
where \(\phi_a\) and \(\phi_b\) are the radial kernels of both quantities. The theoretical \(P_{ab}(z,k)\) is modelled according to the halo-model prescription, adapted to the specifics of the 2MRS catalogue~\cite{Peacock:2000qk,Ando:2017wff}. Finally, we account for the smearing described in \autoref{ssec:gmf} by implementing the following beam:
\begin{equation}\label{eq:beam_har}
    {\cal B}_\ell \simeq \exp\left[-\frac{\ell\,(\ell+1)\,\sigma^2}{2}\right] \,,
\end{equation}
with \(\sigma\) given in \autoref{eq:smear} and assuming \(b=0\).\footnote{The magnetic beam in harmonic space, \autoref{eq:beam_har}, is a rewriting of the Fourier transformation of \autoref{eq:beam}, namely \({\cal B}(k) \deq {\cal F}[{\cal B}(r)] = \exp(-k^2 \sigma^2/2)\) with \({\cal F}\) the Fourier transform operator. In harmonic space this can be approximated using \(k^2 \approx \ell(\ell+1)\) for \(\ell\gtrsim1\).} This corresponds to performing the following substitutions:
\begin{align}
    \cS^\text{CR\,CR}_\ell&\to{\cal B}_\ell^2\,\cS^\text{CR\,CR}_\ell \,,\\
    \cS^\text{g\,CR}_\ell&\to{\cal B}_\ell\,\cS^\text{g\,CR}_\ell \,.\label{eq:Sl_gCR_smeared}
\end{align}

A given UHECR or galaxy observation consist of both a signal, whose harmonic angular power spectrum is given by \autoref{eq:cl_limber}, and a noise power spectrum \(\cN^{ab}_\ell\), which combined give the observed angular power spectrum,
\begin{equation}\label{eq:cell}
    C^{ab}_\ell \deq \cS^{ab}_\ell+\cN^{ab}_\ell \,.
\end{equation}
The noise, being the fields \(\Delta_a(\nv)\) associated to discrete point processes represented by the angular positions of the UHECRs and the galaxies in each sample, is given by
\begin{equation}\label{eq:noise}
	\cN^{ab}_\ell=\frac{\bar{N}_{\Omega,a\wedge b}}{\bar{N}_{\Omega,a}\,\bar{N}_{\Omega,b}} \,,
\end{equation}
where \(\bar{N}_{\Omega,a}\) (\(\bar{N}_{\Omega,b}\)) is the angular number density of points in sample \(a\) or \(b\), and \(\bar{N}_{\Omega,a\wedge b}\) is the angular number density of points shared in common, see~\citep{Urban:2020szk} for details. Given the large number of galaxies in the sample, and the fact that we assume that all UHECRs are generated within any of those galaxies, we shall henceforth neglect the noise term in the cross-correlation power spectrum. This is further justified by the fact that the small angular scales (large \(\ell\) values) are significantly damped by the beam according to \autoref{eq:Sl_gCR_smeared}.

Our goal is to assess how well a full-sky UHECR experiment, with statistics comparable to current facilities, can discriminate among atomic numbers using TSs built from \(C^\text{CR\,CR}_\ell\) and \(C^\text{g\,CR}_\ell\), respectively the harmonic-space UHECR auto-correlation (AC) and UHECR-galaxies cross-correlation (XC) power spectra. By way of example, we consider an admixture of protons (H1) and iron (Fe56), and seek to constrain the fraction of Fe56 \(\fmix=\fFe\), with \(\fFe=0\) being pure proton and \(\fFe=1\) pure iron. This example serves to demonstrate our method, and it is directly generalised to any two atomic numbers \(Z_1\) and \(Z_2\) or combinations thereof. We adopt the following procedure.
\begin{enumerate}
    \item Choose the ``fiducial'' model, which is the model we test against. Our reference model is pure H1 with injection slope \(\gamma=2.3\) and energy cut \(E_\text{cut}=100~\text{EeV}\). For each fiducial model, compute \(N_\text{rea}=250000\) realisations of the expected AC and XC harmonic power spectra.\footnote{We found empirically that with at least 200\,000 realisations the simulated spectra are smooth across all multipoles.} Because we have only a finite number of UHECR events and galaxies, each realisation has a slightly different angular power spectrum, as the signal is scattered by the noise and cosmic variance. Assuming Gaussianity of the \(\Delta^a_{\ell m}\), we can use Wick's theorem to write the harmonic-space power spectrum covariance matrix---which is a four-point correlator---as a sum of all the possible combinations of two-point correlators. Being two-point correlators in harmonic-space the \(\mathcal{S}^{ab}_\ell\) themselves, once we add noise, for each realisation we can draw the \(C^{ab}_\ell\) for every multipole from a Gaussian distribution with a mean which is centred at the AC or the XC harmonic-space power spectra \(\cS^{\text{CR\,CR}}_\ell\) or \(\cS^{\text{g\,CR}}_\ell\), and a standard deviation of:
    \begin{align}
    \sigma_\text{AC}&=\sqrt{\frac{2}{2\ell+1}}C^{\text{CR\,CR}}_\ell \,,\label{eq:StdforHistsAC}\\
    \sigma_\text{XC}&=\sqrt{\frac{\left({\cS^{\text{g\,CR}}_\ell}\right)^2+C^{\text{g\,g}}_\ell\, C^{\text{CR\,CR}}_\ell}{2\ell+1}} \,,\label{eq:StdforHistXC}
    \end{align}
    for the AC and the XC, respectively. Notice that, although we draw independently for the harmonic AC and XC power spectra, the two are in reality correlated. We have verified that the, numerically much slower, full multi-variate draw makes no appreciable difference.
    \item For each realisation, compute the TSs to obtain their statistical distributions. The TSs we choose to compute are: the dipole \(C^{ab}_1\); the quadrupole \(C^{ab}_2\); the octupole \(C^{ab}_3\); the total harmonic angular power for all multipoles \(\ell\in[1,1000]\), that is
    \begin{align}\label{eq:c_def}
        C^{ab}\deq\sum_\ell C^{ab}_\ell \,.
    \end{align}
    As we will see the most powerful among them is the total power \(C^{ab}\), so we will focus on this quantity in what follows, and leave the discussion of the individual multipoles for the Appendix. Because we are using a \(1\,\deg\) resolution beam to account for the fact that UHECR experiments cannot resolve angular scales smaller than that, including multipoles above 1000 would make no difference; moreover, it was found in~\cite{Urban:2020szk} that for \(\ell\gg100\) the power spectra are well below the shot noise due to the magnetic field suppression.
    \item For a given value of the ``test'' model parameter, in our case \(\fmix\) (more specifically \(\fFe\)), generate the synthetic AC and XC power spectra in the same way, and compute the values of the same TSs.
    \item Compute at which confidence level (CL) \(q\), quantified with respect to the test TS distribution, a single realisation of the fiducial model is incompatible with the test model---this is done fitting \(N_\text{test}=250000\) realisations of the test model to a Gaussian. The CL computed in this way represents the confidence with which one particular instance of a hypothetical experiment as drawn from the fiducial distribution, is able to reject the test model, given the specified experimental setup (e.g.\ the number of observed events).
    \item Repeat this step for all the \(N_\text{rea}\) realisations of the fiducial model, in order to obtain the percentage \(n\) of experiments that, given a value of \(\fmix\), can exclude the test model against the fiducial at at least \(q\) CL.
    \item Repeat the entire procedure for varying model parameter \(\fmix\). All together, this gives the percentage \(n(\fmix;q)\) of experiments which will be able to exclude \(\fmix\) at at least \(q\) CL, or---this information can be read across \(q\) as well as across \(\fmix\)---exclude \(\fmix\) or larger (when the fiducial is \(\fmix=0\), otherwise \(\fmix\) or smaller if the fiducial is \(\fmix=1\)) at \(q\) CL. Notice that the reasoning can be reversed, and the quantity \(\tilde{n}(\fmix;q)\deq1-n(\fmix;q)\) gives the percentage of experiments for which the test model labelled by \(\fmix\) is compatible with the fiducial (that is, experiments which cannot distinguish the two models) within \(q\) CL, or the percentage of experiments for which \(\fmix\) or smaller (when the fiducial is \(\fmix=0\), otherwise \(\fmix\) or larger) is compatible with the fiducial at \(q\) CL.
\end{enumerate}

\begin{figure}[htbp]
    \centering
    \includegraphics[width=1.0\columnwidth]{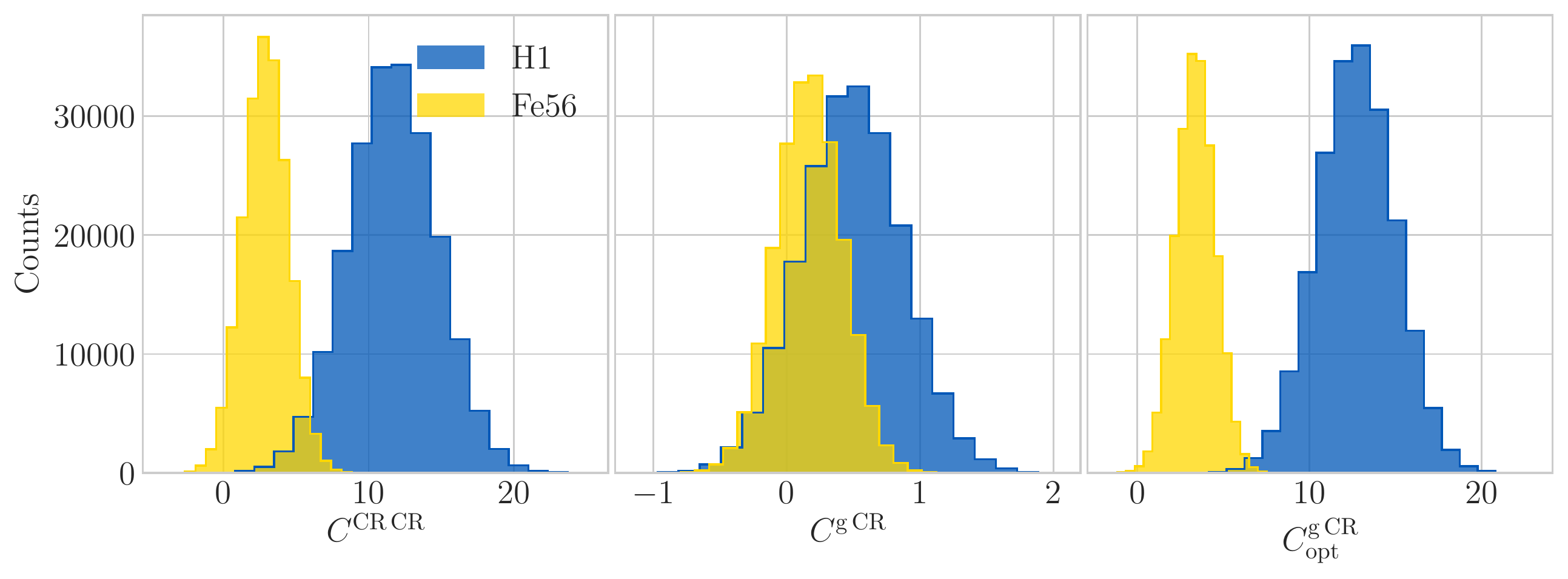}
    \caption{Probability distribution of the \(N_\text{rea}=250\,000\) realisations of the TS \(C^{ab}\) for the AC (\(a=\text{CR}=b\), left panel), the XC (\(a=\text{g}\), \(b=\text{CR}\), middle panel), and XC with the optimal weights of \autoref{eq:opt_w} (right panel). The blue histograms are for the case of \(Z=1\) (H1) injection, whereas the yellow ones are for \(Z=26\) (Fe56) injection.}
    \label{fig:AC_hist}
\end{figure}

In Figs.~\ref{fig:AC_hist}, for our benchmark model with \(\Ecut=100~\text{EeV}\) we show the histograms obtained according to the method just described, respectively for the AC, XC, and XC\(_\text{opt}\) (wherein the optimal weights of \autoref{eq:opt_w} have been applied). Notice how, as expected, the XC performs more poorly than the AC (that is, the histograms for H1 and Fe56 overlap more and are thus harder to be told apart); the XC however becomes competitive once optimal weights are applied to the galaxy catalogue. In what follows we will always work with the optimal weights, as is standard practice in the literature (albeit typically not formulated in these terms), see e.g.\ \cite{Koers:2008ba}.

\begin{figure}[htbp]
    \centering
    \includegraphics[width=1.0\columnwidth]{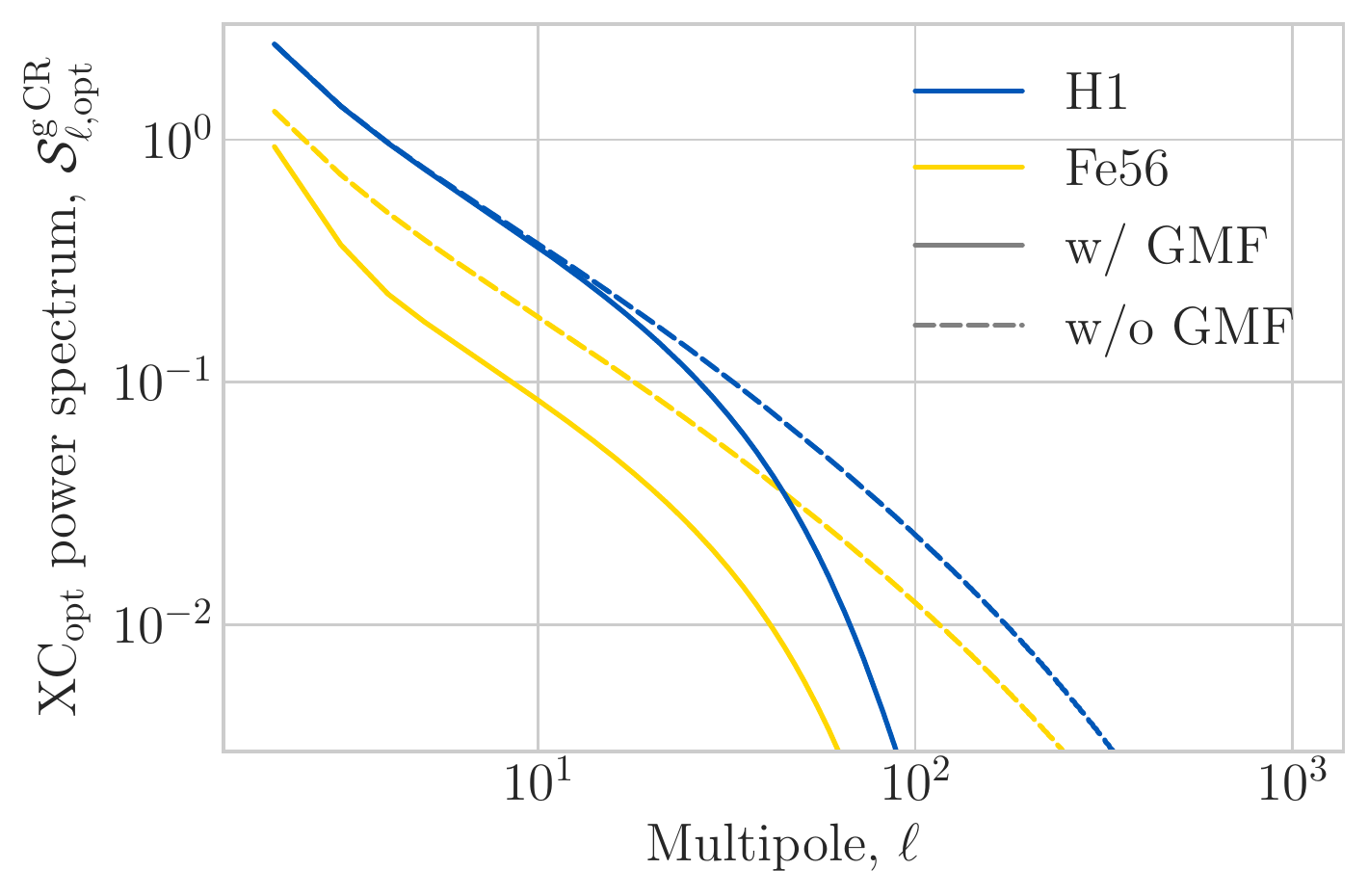}
    \caption{Optimal angular XC signal \(\cS^\text{g\,CR}_{\ell,\text{opt}}\) for H1 (blue lines) and Fe56 (yellow lines) injection, with (solid lines) and without (dashed lines) GMF deflections accounted for.}
    \label{fig:XC_opt_spectra}
\end{figure}

Notice that although the histograms are both Gaussian, they peak at different values and also have non-equal widths depending on the atomic number. This happens due to the fact that the harmonic power spectra for H1 and Fe56 are different: H1 has more power than Fe56 because it is much less affected by the GMF deflections, while having a comparable propagation horizon.\footnote{This is considerably different than other primaries such as silicium or oxygen, which attenuate much more We leave the study of such cases for future work \citep[see][]{upcoming}.} In \autoref{fig:XC_opt_spectra} we show the H1 (blue) and the Fe56 (yellow) power spectra with (solid) and without (dotted) the suppression caused by the GMF. Therefore, because the standard deviations for the two spectra are proportional to the means, that is the signals (see \autoref{eq:StdforHistsAC} and \autoref{eq:StdforHistXC}), the TS values \(C^{ab}\) for the H1 case have a larger scatter than those of Fe56. These differences in means and standard deviations consistently imply that this method is not symmetric under the exchange of fiducial and test models.

\section{Results}\label{sec:results}

\begin{figure}[htbp]
    \centering
    \includegraphics[width=1.0\textwidth]{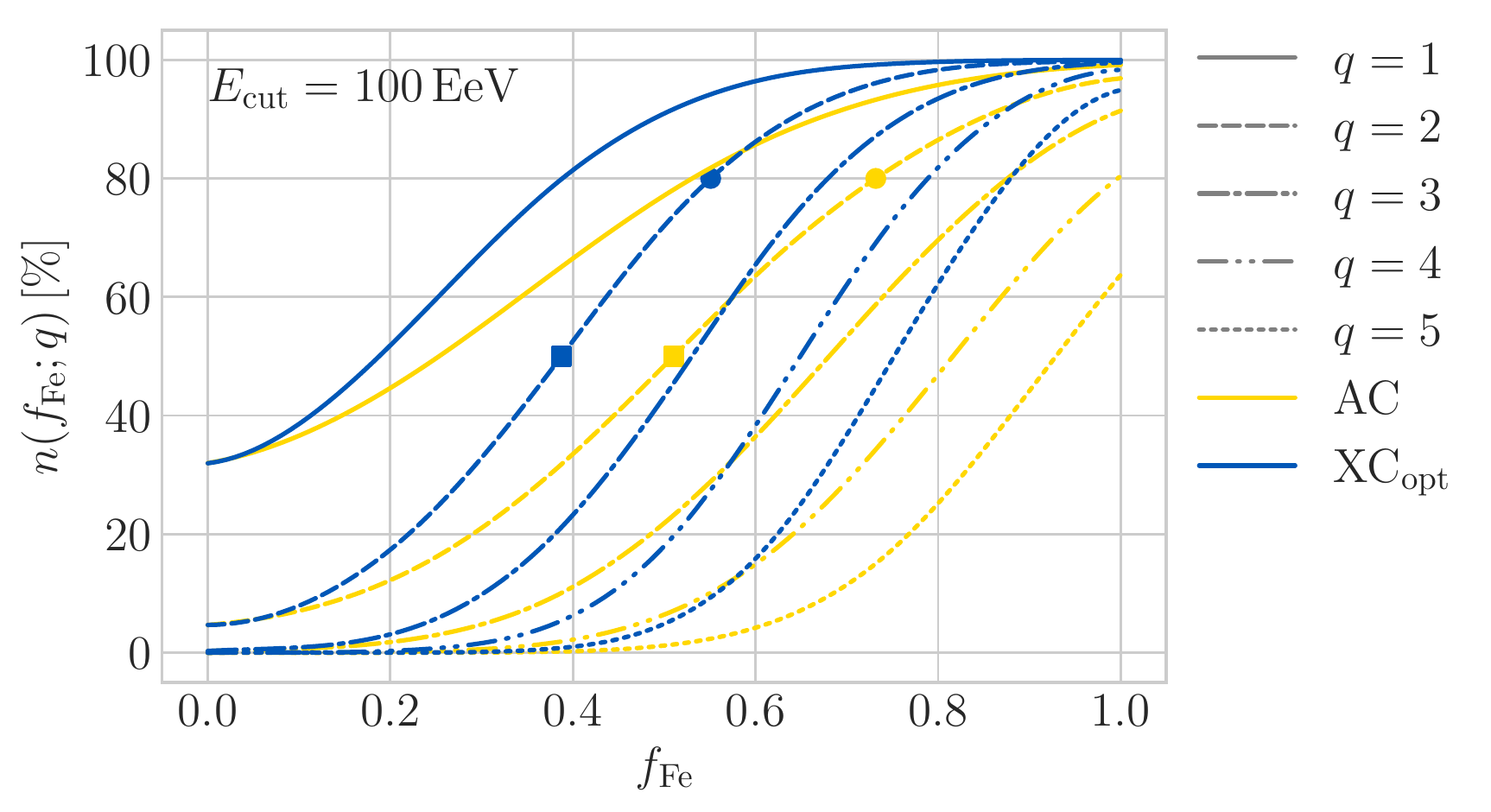}
    \caption{Percentage of experiments \(n(\fFe;q)\) that will be able to exclude \(\fFe\) or more at \(q\) CL for different choices of \(q\): from \(q=1\) (solid) to \(q=5\) (dotted), for our benchmark model with \(\Ecut=100~\text{EeV}\) and \(Z=1\). The TSs used are the total angular AC power \(C^\text{CR\,CR}\) in yellow, and the total optimal angular XC power \(C^\text{g\,CR}_\text{opt}\) in blue---notice the different colour coding w.r.t.\ \autoref{fig:AC_hist} and \autoref{fig:XC_opt_spectra}. Circle and square markers show the \(\fFe\) values respectively corresponding to \(n=50\%\) and \(80\%\) (yellow for AC and blue for XC).}
    \label{fig:n_f_100_H1}
\end{figure}

To assess the power of the total angular power \(C^{ab}\) to exclude a mixed proton-iron composition when the fiducial model is pure protons, in \autoref{fig:n_f_100_H1} we plot the percentage of experiments \(n(\fFe;q)\) that will be able to exclude a fraction of iron equal to \(\fFe\) or higher at \(q\) CL (for different choices of \(q\)), for our benchmark model with \(\Ecut=100~\text{EeV}\) and \(Z=1\). The fiducial is recovered by setting \(\fFe=0\). The yellow curves are for the AC while the blue curves are for the XC\(_\text{opt}\), in which the optimal weights of \autoref{eq:opt_w} have been applied. We see that in this case the optimised XC\(_\text{opt}\) performs better for nearly all values of \(\fFe\). This is because the XC is more sensitive to magnetic deflections than the AC---even though the magnetic beam appears twice in the AC and only once in the XC, the XC power is at much smaller scales than that of the AC, and small scales are much more influenced by magnetic deflections---so it is more informative about them. As a consistency check, notice that at \(\fFe=0\) the \(q=1\) (\(q=2\)) line gives approximately \(n=32\) (\(n=5\)), that is, a ``probability'' of excluding the test model, which in this case is the fiducial itself, corresponding to \(32\%\) (\(5\%\)) as expected.

\begin{figure}[htbp]
    \centering
    \includegraphics[width=1.0\columnwidth]{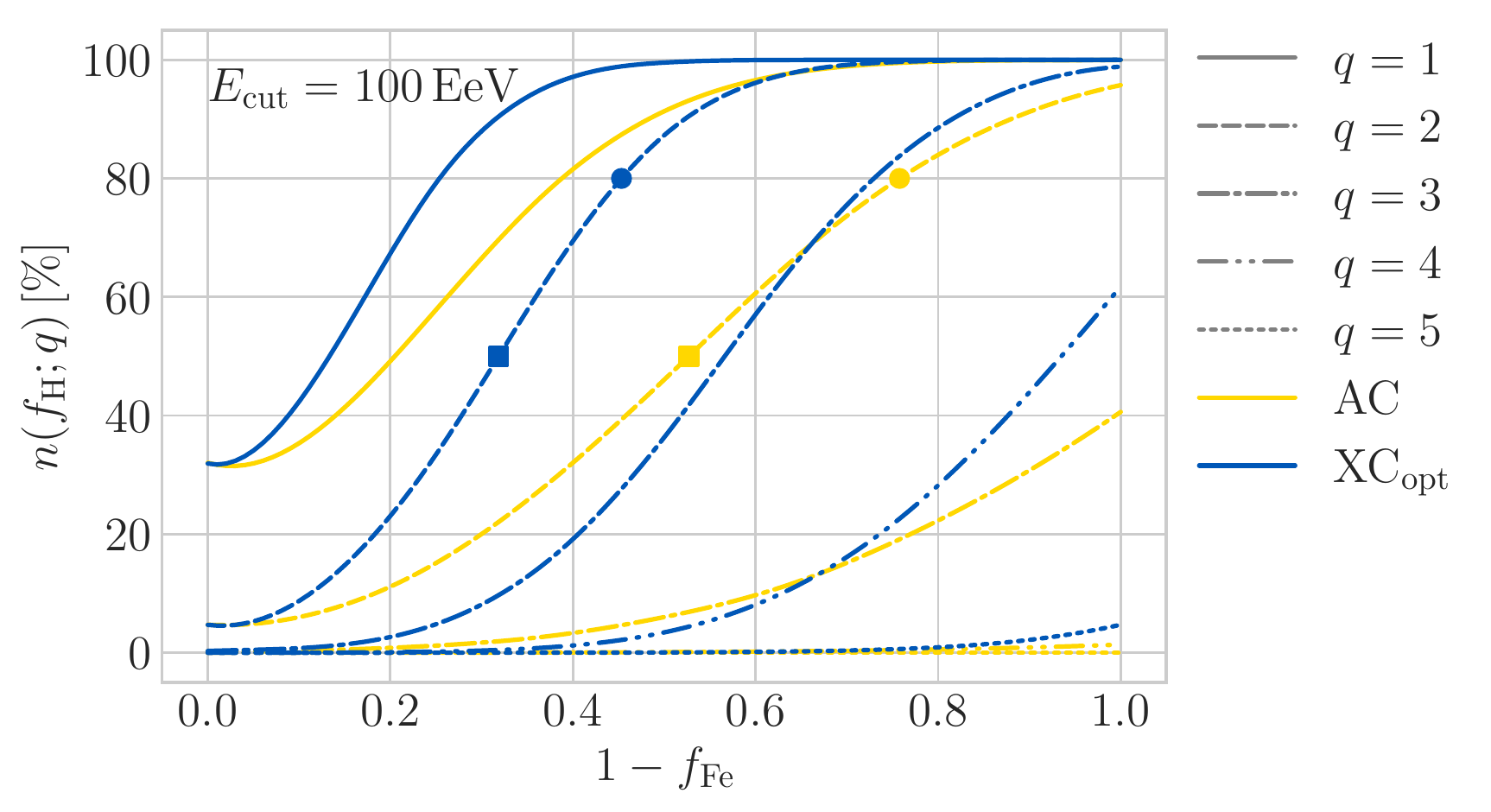}
    \caption{Same as \autoref{fig:n_f_100_H1} but exchanging fiducial and test model: in this case we test a fraction \(\fH\deq1-\fFe\) of \(Z=1\) injection against a \(Z=26\) fiducial.}
    \label{fig:n_f_100_Fe56}
\end{figure}

In order to test the asymmetry between fiducial and test model, in \autoref{fig:n_f_100_Fe56} we plot the percentage of experiments \(n(\fH;q)\), where \(\fH\deq1-\fFe\) is the fraction of H1 using a pure Fe56 fiducial model, that will be able to exclude \(\fH\) or more at \(q\) CL (for different choices of \(q\)), for energy cut \(\Ecut=100~\text{EeV}\). Notice that the fiducial in this case is recovered for \(\fH=0\) (\(\fFe=1\)). As expected, because the distribution of the TS in the case of Fe56 is narrower, when the distance is counted in Fe56 standard deviations, it is more likely that a test realisation of H1 is ``further'' away from the bulk of the narrower Fe56 distribution than vice versa.

\begin{figure}[htbp]
    \centering
    \includegraphics[width=1.0\columnwidth]{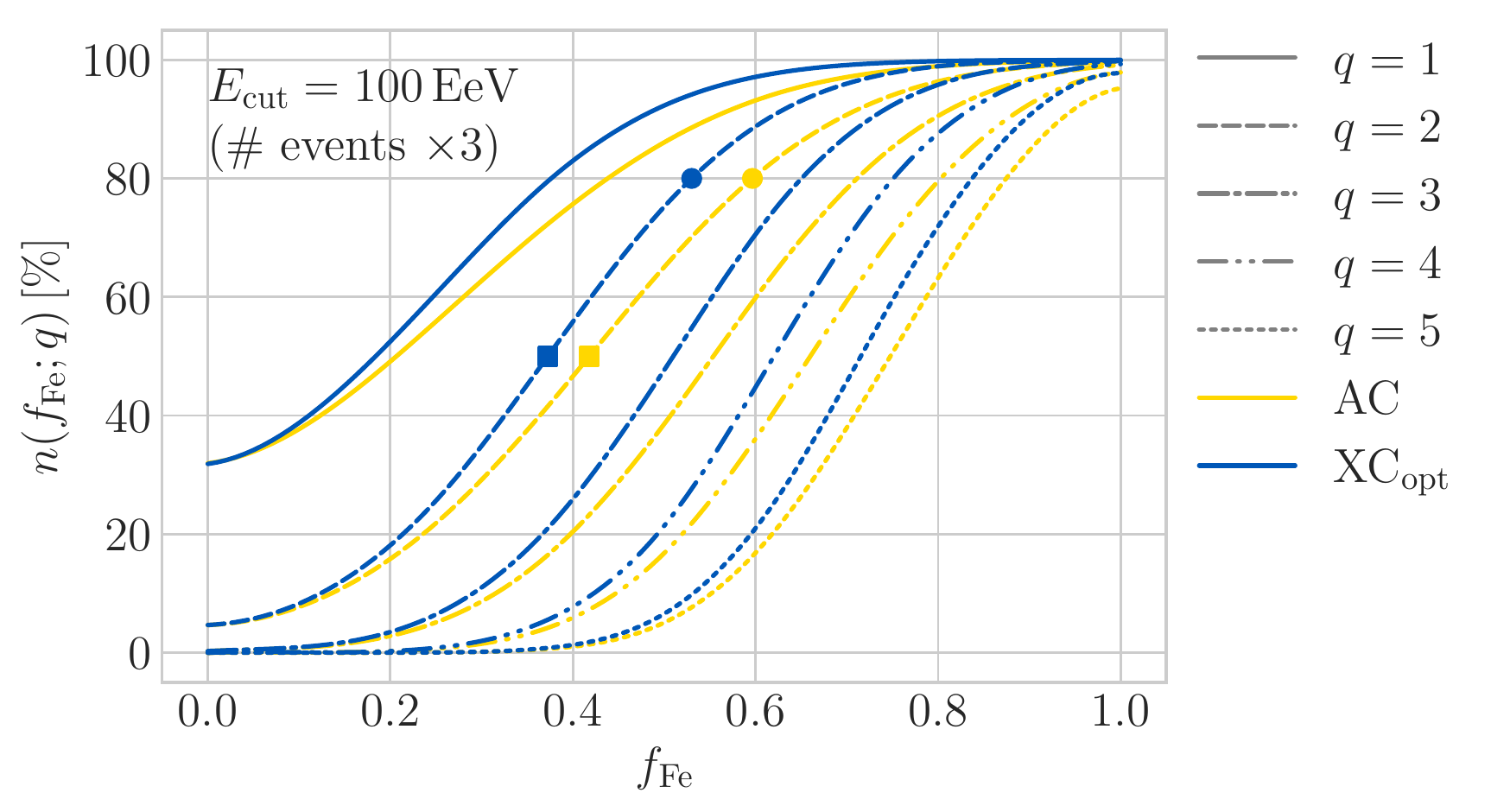}
    \caption{Same as \autoref{fig:n_f_100_H1} but three times more events, that is, \(\nev\approx100\) for \(\Ecut=100~\text{EeV}\).}
    \label{fig:n_f_100_H1_3x}
\end{figure}

In order to forecast the constraining power with future experimental facilities, in \autoref{fig:n_f_100_H1_3x} we plot the percentage of experiments \(n(\fFe;q)\) that will be able to exclude \(\fFe\) or more at \(q\) CL (for different choices of \(q\)), for energy cut \(\Ecut=100~\text{EeV}\) but with three times more events, namely \(\nev\approx100\) for \(\Ecut=100~\text{EeV}\). The improvement is significant for both the AC and the XC\(_\text{opt}\), but it is especially so for the AC.  This is expected because increasing the number of events reduces the shot noise of the AC much more pronounced than for the XC, which relies mostly on the much larger number of galaxies to trace the distribution of sources.  Nonetheless, for \(\nev\approx100\) at \(\Ecut=100~\text{EeV}\), the AC remains slightly less informative than the XC\(_\text{opt}\).

If we modify the injection spectrum spectral index or by introducing a high-energy cutoff, the main change is brought about by the different child proton fraction of Fe56. For instance, changing the spectral indices to \(\gamma=2.1\) and \(\gamma=2.5\) gives \(f_\text{p}\approx0.40\) and \(f_\text{p}\approx0.12\), respectively. In the first case the fraction of Fe56 which we can constrain at \(2\sigma\) CL degrades to about \(\fFe\approx0.5\) (0.7) in 50\% (80\%) of the imagined experiments; in the latter case there is an improvement of about 10\% compared to the \(\gamma=2.3\) case, as expected, because with fewer child protons the Fe56 flux is less contaminated. If the injection spectrum turns out to be much harder, viz.\ \(\gamma<2\), the number of child protons at the highest energies will begin to outnumber the heavy nuclei arriving on Earth, making Fe56 and H1 injection indistinguishable. If we introduce a high-energy cutoff in the spectrum the amount of child protons that come from the disintegration of the highest-energy Fe56 nuclei will drop, therefore improving our results. In the most extreme case, when the cutoff is around \(A=56\) times \(E_\text{cut}\), there would be no child protons at all, and the \(\fFe\) we can constrain drops to approximately 0.3 (0.4) in 50\% (80\%) of the measurements.

\section{Discussion and outlook}\label{sec:conclusions}

In this paper we have discussed a new method to test different UHECR injections models, in particular to differentiate between different atomic numbers \(Z\), using the harmonic angular UHECR AC and UHECR-galaxies XC power spectra. As a concrete example we have assessed how well a full-sky UHECR experiment with statistics comparable to current facilities can discriminate between different admixtures of H1 (\(Z=1\)) and Fe56 (\(Z=26\)) injected nuclei, both with injection slope \(\gamma=2.3\). Our results show that the best observable for this task is the total angular power \(C^{ab}=\sum_\ell C^{ab}_\ell\) for either the AC or XC. Moreover, when optimal weights that take into account the propagation properties of UHECRs are applied to the galaxy catalogue, the XC slightly outperforms the AC in most tests. We did not find a significant dependence on energy in the range \(40~\text{EeV}\leq\Ecut\leq100~\text{EeV}\), indicating that, within our modelling of the GMF effects, the effects of the GMF deflections at different energies are roughly compensated by the different UHECR propagation horizon and number of events. Quantitatively, we find that about \(50\%\) (\(80\%\)) of experiments measuring the XC \(C^\text{g\,CR}_\text{opt}\) with \(\Ecut=100~\text{EeV}\) would be able to exclude \(\fFe\gtrsim0.39\) (\(\fFe\gtrsim0.55\)) against a \(Z=1\) fiducial at \(q=2\) CL (that is, at \(2\sigma\); in the case of the AC \(C^\text{CR\,CR}_\text{opt}\), these numbers become \(\fFe\gtrsim0.51\) (\(\fFe\gtrsim0.73\) ) ).

The method we described here is similar to what was pushed forward in~\cite{Kuznetsov:2020hso}; the main difference is that our TS is the total angular harmonic AC and XC power spectra encoded in \(C^{ab}\) instead of an individual averaged angle quantifying the deflections in the GMF. While our approximations and assumptions are different from~\cite{Kuznetsov:2020hso}, we find qualitative agreement in that it is easier to constrain a small amount of Fe56 on an H1 map than vice versa.  Moreover, because the angular harmonic AC is much less sensitive to the details of the GMF structure~\cite{Tinyakov:2014fwa}, as is the XC~\cite{Urban:2020szk}, we also expect that our method remains informative even in the face of an uncertain knowledge of the GMF parameters.  This is more so because we use the \emph{total} harmonic angular power \(C^{ab}\), whereas, as shown in \autoref{app:some}, individual multipoles are not as constraining.

Our results imply that already with current data significant quantitative statements about the composition of UHECRs at the highest energies can be made. Moreover, in the near future with the completion of the expansions of both the Telescope Array detector TAx4~\cite{Abbasi:2021hO}, and in the longer term with the advent of next-generation detectors such as GRAND~\cite{Alvarez-Muniz:2018bhp} and POEMMA~\cite{Olinto:2017xbi}, the larger available data sets would quickly improve the sensitivity of our method and allow to determine (or exclude) with more precision several composition models.

In the future, in order to further understand the capability of our method in determining the injection properties of the observed UHECR events, we plan to extend our analysis in three ways. Firstly, we will test several other injection models, for example oxygen (\(Z=8\)) or silicium (\(Z=14\)), which have a much shorter propagation horizon and are less sensitive to GMF deflections compared to Fe56. Secondly, to assess how stable this method is with respect to changes in the structure of the GMF, we plan to extend our GMF model by including the effects of the large-scale GMF, a task which will require a full simulation of the propagation of UHECRs in the Milky Way; moreover, even the small-scale GMF model can be refined by accounting for the fact that the deflections are latitude-dependent (see \autoref{eq:smear}). Thirdly, we plan to construct a test that makes use of the full information encoded in the harmonic-space power spectra \(C^{ab}_\ell\) instead of compressing them into one single number (i.e.\ the total power \(C^{ab}\)). This will permit a much more precise determination of the parameters of the injection model and of the GMF because the AC and XC at each multipole \(\ell\) is an independent observable and is affected differently by changes in \(Z\) or \(B\). We further speculate that, with the next-generation of experiments, a refined and uncompressed version of our method could fully disentangle the degeneracy between the atomic number \(Z\) and the GMF field strength and allow a precise study of the GMF properties with UHECR data.

\acknowledgments
FU wishes to thank M.~Kuznetsov for useful correspondence and A.~Bakalova for feedback on the draft. KT and FU are supported by the European Regional Development Fund (ESIF/ERDF) and the Czech Ministry of Education, Youth and Sports (MEYS) through Project CoGraDS~-~\verb|CZ.02.1.01/0.0/0.0/15_003/0000437|. SC acknowledges the `Departments of Excellence 2018-2022' Grant awarded by the Italian Ministry of University and Research (\textsc{mur}, L.\ 232/2016).

\appendix
\counterwithin{figure}{section}

\section{Lower energy cuts and individual multipoles}\label{app:some}

\begin{figure}[htbp]
    \centering
    \includegraphics[width=1.0\columnwidth]{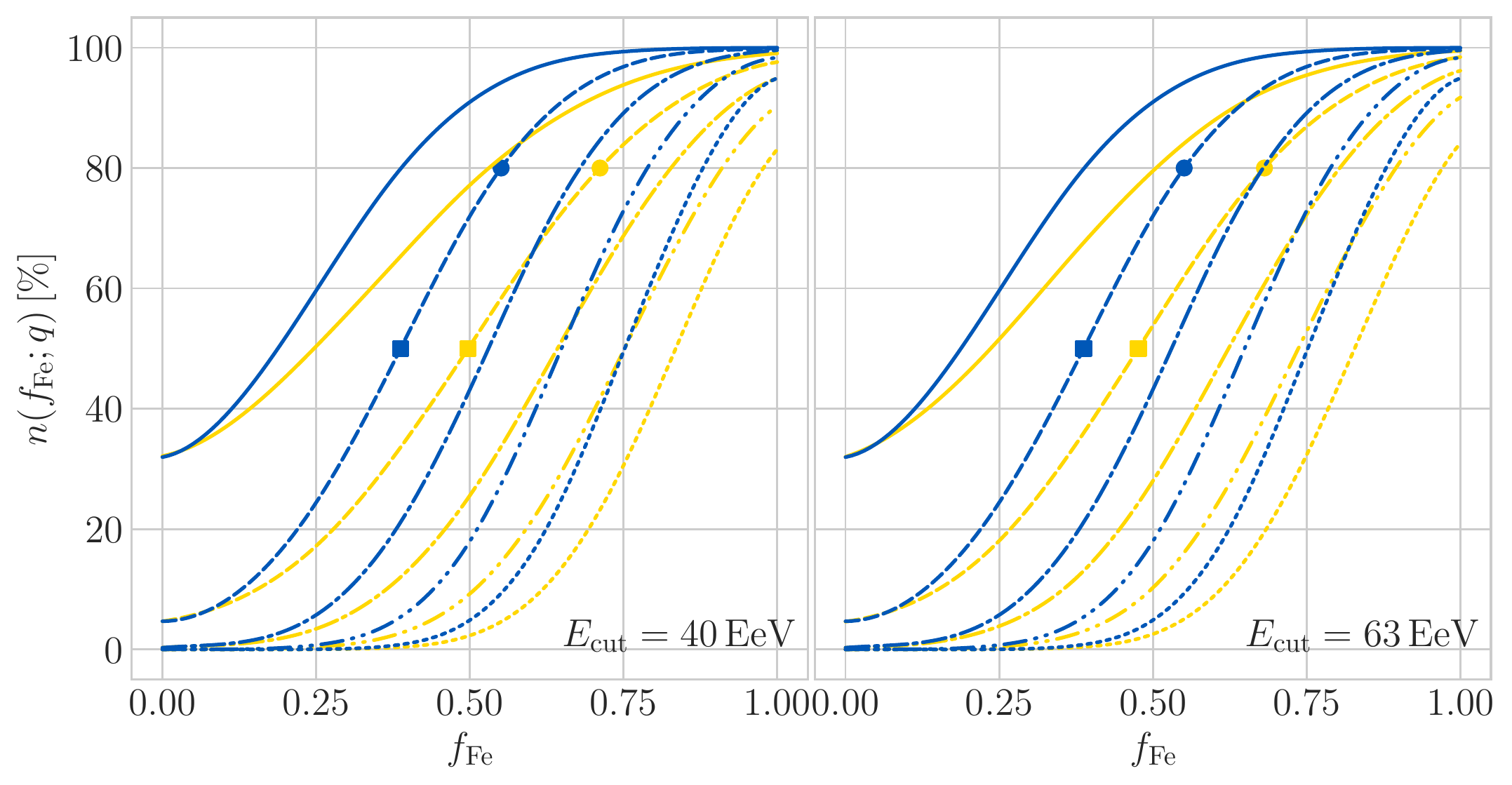}
    \caption{Same as \autoref{fig:n_f_100_H1} but for energy cuts \(\Ecut=40~\text{EeV}\) (left) and \(\Ecut=63~\text{EeV}\) (right).}
    \label{fig:n_f_40+63_H1}
\end{figure}

For completeness, in \autoref{fig:n_f_40+63_H1} we plot the percentage of experiments \(n(\fFe;q)\) that will be able to exclude \(\fFe\) or more at \(q\) CL (for different choices of \(q\)), for energy cuts \(\Ecut=40~\text{EeV}\) (left panel) and \(\Ecut=63~\text{EeV}\) (right panel). Comparing these results with the \(\Ecut=100~\text{EeV}\) ones in \autoref{fig:n_f_100_H1} we see that the latter performs better, although the differences are marginal. In other words, we do not find a strong dependence on \(\Ecut\) within our model for the GMF effects.  Notice that we do not model any energy-dependence at injection, that is, the atomic number \(Z\) does not depend on energy.

\begin{figure}[htbp]
    \centering
    \includegraphics[width=1.0\columnwidth]{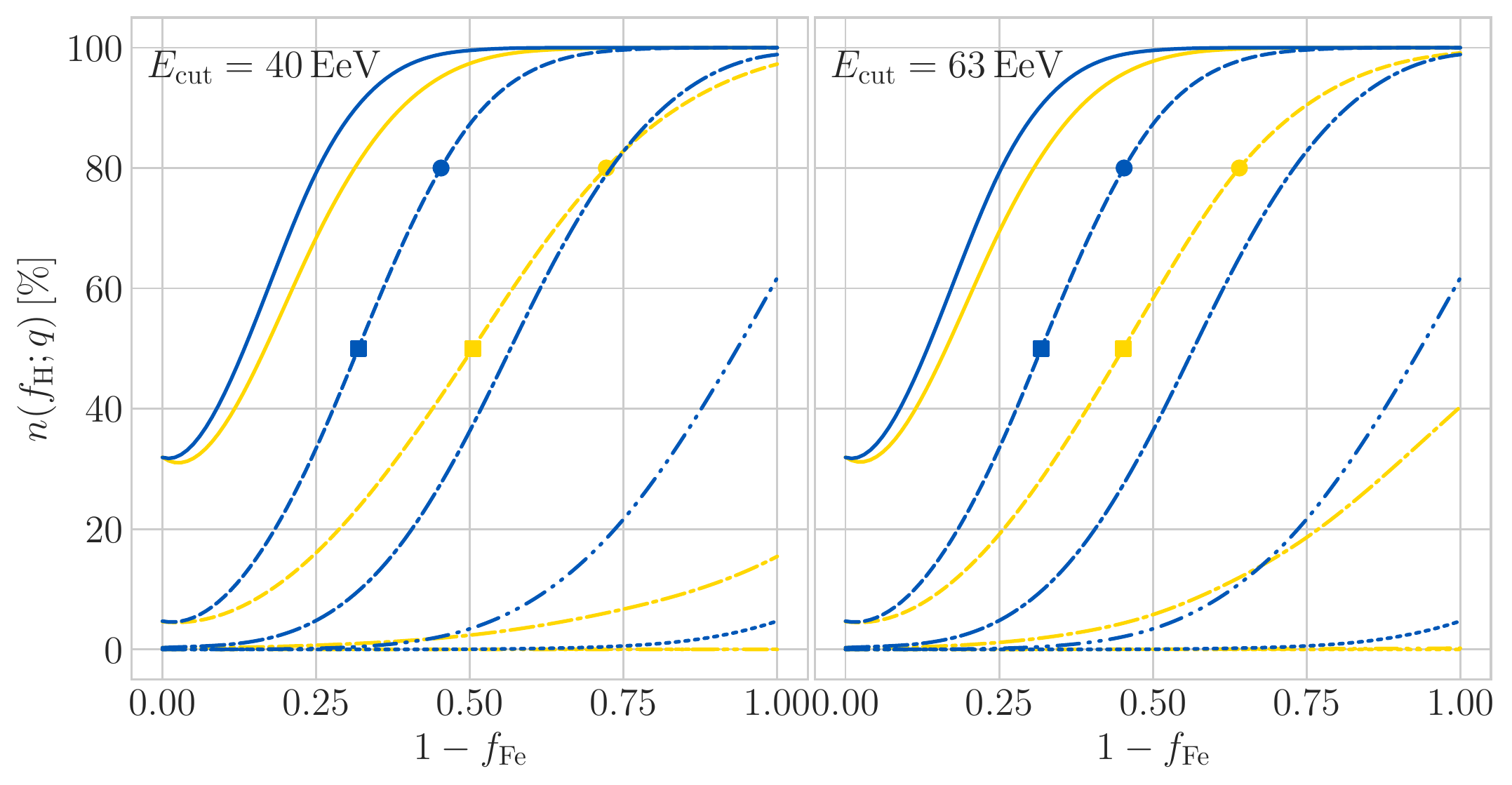}
    \caption{Same as \autoref{fig:n_f_100_H1} but for energy cuts \(\Ecut=40~\text{EeV}\) (left) and \(\Ecut=63~\text{EeV}\) (right) and exchanging fiducial and test model.}
    \label{fig:n_f_40+63_Fe56}
\end{figure}

Then, in \autoref{fig:n_f_40+63_Fe56} we plot the percentage of experiments \(n(\fH;q)\), where \(\fH=1-\fFe\) is the fraction of H1 against a Fe56 fiducial, that will be able to exclude \(\fH\) or more at \(q\) CL (for different choices of \(q\)), for energy cuts \(\Ecut=40~\text{EeV}\) (left) and \(\Ecut=63~\text{EeV}\) (right). The performances of the TS are similar (albeit slighly worse) than in the case of \(\Ecut=100~\text{EeV}\) shown in \autoref{fig:n_f_100_Fe56}. Once again we do not observe a significant dependence on the energy cut.

\begin{figure}[htbp]
    \centering
    \includegraphics[width=1.0\columnwidth]{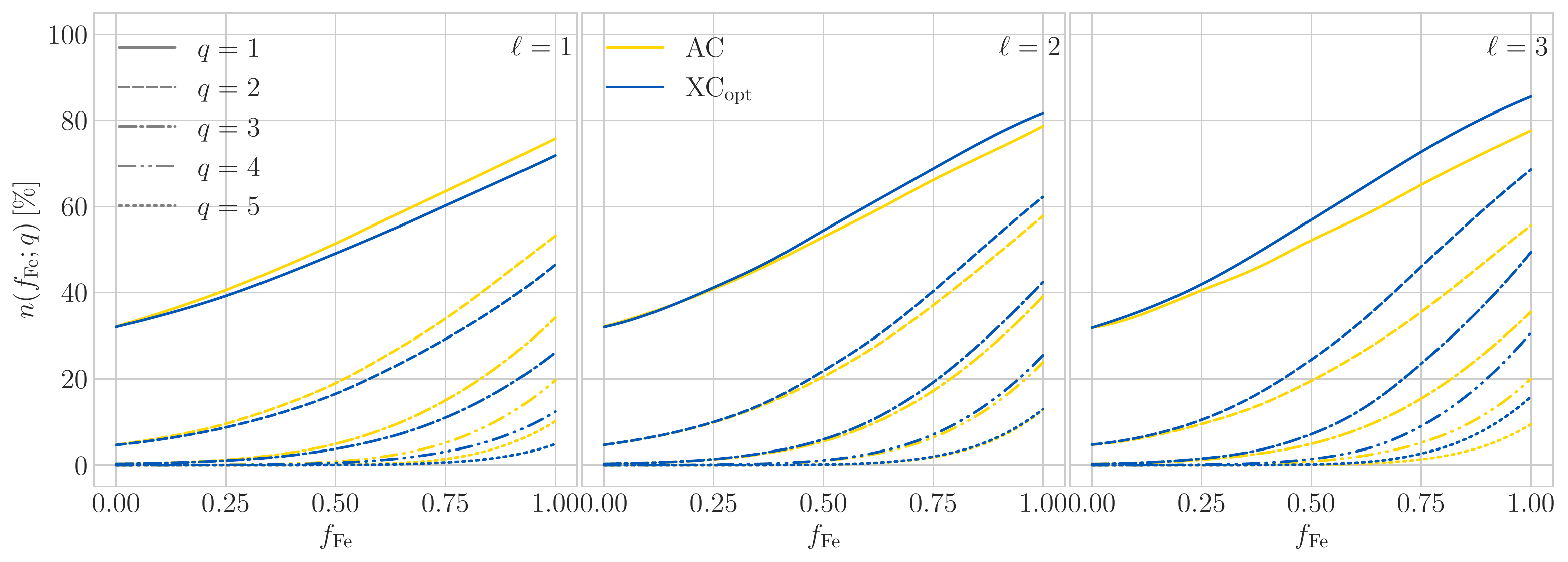}
    \caption{Same as \autoref{fig:n_f_100_H1} but using individual multipoles as TS: dipole \(\ell=1\) (left panel), quadrupole \(\ell=2\) (central panel) and octupole \(\ell=3\) (right panel).}
    \label{fig:n_f_100_ell_Fe56}
\end{figure}

Lastly, in \autoref{fig:n_f_100_ell_Fe56} we show the performance of the other TS (other than the total angular power \(C^{ab}\)), namely the dipole \(\ell=1\), quadrupole \(\ell=2\) and octupole \(\ell=3\), in the left, central and right panels, respectively. These individual multipoles do not have the same constraining power as the total angular power (as expected), although one can imagine situations in which, if they could be detected, they would provide useful information about both the UHECR composition and the GMF model which the total angular power \(C^{ab}\) would conceal.

\section{The effect of varying the displacement angle}\label{app:deflect}

\begin{figure}[htbp]
    \centering
    \includegraphics[width=1.0\columnwidth]{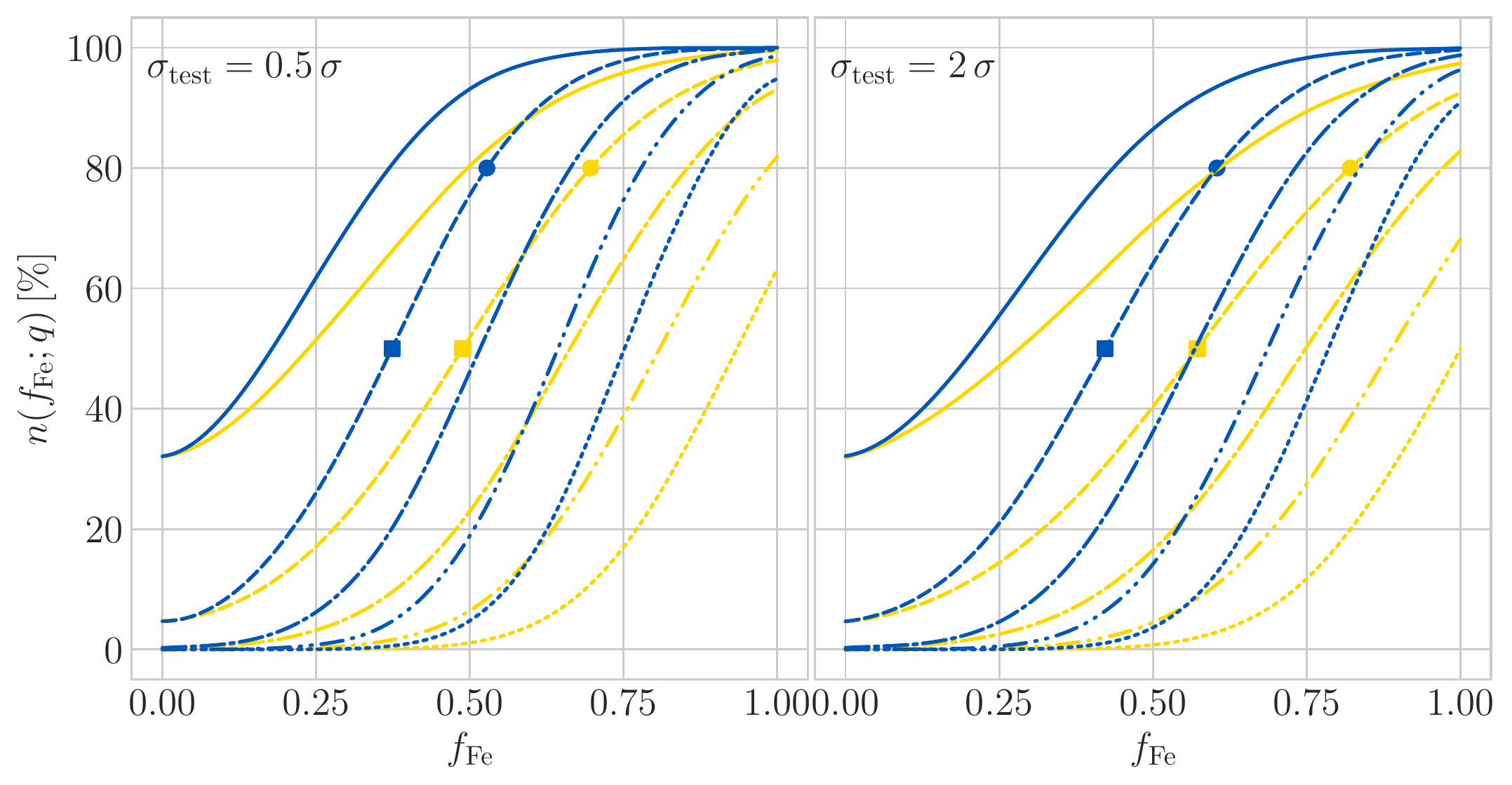}
    \caption{Same as \autoref{fig:n_f_100_H1} but for deflections of half (left) and double (right) the reference values given by \autoref{eq:smear}.}
    \label{fig:n_f_smear}
\end{figure}

To assess the impact of our GMF modelling in our results, in \autoref{fig:n_f_smear} we plot once again the percentage of experiments \(n(\fFe;q)\) that will be able to exclude \(\fFe\) or more at \(q\) CL (for different choices of \(q\)), for energy cut \(\Ecut=100~\text{EeV}\) with deflections quantified as \(\sigma_\text{test} = 0.5\,\sigma\) and \(\sigma_\text{test} = 2\,\sigma\) (left and right panel, respectively). We remind the reader that \(\sigma\) is given by \autoref{eq:smear}. Qualitatively, the outcome remains the same, with the XC outperforming the AC in discriminating Fe56 and H1, but with the strength of the constraints degrading/improving by a few percent with larger smaller/deflections.

\bibliography{references}

\end{document}